\documentclass[12pt]{article}

\usepackage[margin=1in]{geometry}
\usepackage{setspace}
\usepackage{amsmath,amssymb}
\usepackage{graphicx}
\usepackage{physics}
\usepackage{cite}
\usepackage{hyperref}
\usepackage{fancyhdr}
\usepackage{titlesec}

\usepackage{mathrsfs}
\usepackage[normalem]{ulem}
\usepackage{caption}

\titleformat{\section}
{\large\bfseries}
{\thesection.}{0.5em}{}

\title{\textbf{Title of the Essay}}

\date{\today}

\begin{document}


\begin{titlepage}

\centering

{\Large \textbf{On the nature of entangling photons in 
horizon-induced decoherence}}

\vspace{1cm}

Max Joseph Fahn$^{1,2}$\footnote{maxjoseph.fahn@unibo.it}, 
Alessandro Pesci$^{2}$\footnote{pesci@bo.infn.it}

\vspace{0.5cm}

$^{1}$Dipartimento di Fisica e Astronomia, Università di Bologna, 
Via Irnerio 46, 40126 Bologna, Italy \\
$^{2}$INFN, Sezione di Bologna, 
Via Irnerio 46, 40126 Bologna, Italy \\

\vspace{1cm}


\vspace{1cm}

\textbf{Abstract}

\begin{quote}
Recently, it was discussed how
the presence of a Killing
horizon induces decoherence on a
quantum system in a superposition
of states. Focusing on the
case of an electrically-charged system
with superposed positions, this would
happen due to ``entangling'' photons
crossing the horizon while carrying
information on the superposition. Purpose
of this essay is to
investigate this process in connection
with black hole thermodynamics and
the ensuing entropy bounds. We
show that an apparent tension
arising with the latter is
resolved provided the entangling
photons, expressing a modification of the field
at, as well as inside
the horizon, do not give
rise to a flux across
it. The storage of information
in this field, not retrievable
from an outside observer, causes
the superposition to decohere.
\end{quote}

\vspace{0.5cm}

\textit{Essay written for the Gravity Research Foundation 2026 Awards for Essays on Gravitation. It received an Honorable Mention.}

\vfill

\end{titlepage}


\section{Introduction and context}
Recent results have suggested
that black holes
induce decoherence on a system in a superposition
of states \cite{Danielson:2022tdw,Danielson:2023sga}.
It has been shown indeed that,
even in case the system undergoes
almost completely adiabatic transformations,
some decoherence
appears, arising from the mere presence of a horizon,
with decoherence functional
growing linearly with the time during which the superposition exists.

This has been confirmed in a variety of approaches
and systems, being the latter neutral or charged
(with scalar or electromagnetic charge) and being the phenomenon described
globally (for the whole of spacetime, see e.g. \cite{Danielson:2022tdw,Danielson:2023sga,Gralla:2023oya})
or locally, from the vantage point of a local observer where the quantum system is (see for instance \cite{Wilson-Gerow:2024ljx,Danielson:2024yru});
moreover, this has been shown to hold in all generality,
to wit, for generic Killing horizons,
including the Rindler case \cite{Danielson:2023sga}.

The considered circumstances
are the transposition to the vicinity of a horizon
of a much studied gedanken experiment (see \cite{Belenchia:2018szb} and references therein)
encompassing a (charged or neutral) microparticle $A$ put by Alice
in a superposition of locations,
whose (electromagnetic or gravitational) field
is probed through another microparticle $B$,
free to move in a coherent manner some distance $D$ apart,
and on which Bob performs
measurements of position.

If the sourced field is quantum, $B$ undergoes a superposition
of time evolutions,
and then, if a measurement of position of $B$ allows Bob to discriminate
the branch of $A$'s superposition,
by the complementarity principle (no possible interference pattern if
the branch is known), the superposition state of $A$ must decohere
as can be checked by Alice on recombining her particle.
On the other hand, if no measurement is done by Bob,
then, if Alice carefully recombines her particle,
the time evolutions of $B$ also recombine smoothly,
and $A$'s (and $B$'s) coherence is preserved.

So far so good, but some tension with the causality principle
might arise when both the experimenters, Alice and Bob,
are able to do their operations in a time smaller than the distance time
between $A$ and $B$ so that they are causally disconnected;
in this case indeed:
Is Alice going to obtain interference fringes in spite of Bob
having determined the branch?
Or, inversely, how is it possible that her particle decoheres without any notice
of Bob's accomplishments?

A brilliant way out of the apparent conundrum has been proposed \cite{Belenchia:2018szb}:
Focusing on the electromagnetic case (but a similar solution
can be invoked for a linearised quantum gravitational field),
when circumstances are such that
(read, when charge $q$ 
and 
separation $d$
are large enough that)
Bob can do which-path discrimination in a time smaller
than the distance time,
then, for that same circumstances, if Alice recombines
in a similarly quick time, the particle necessarily emits
a photon that carries away phase information and the coherence of the charge $A$ is lost.

However, performing the same experiment near a black hole with Bob
possibly hiding behind the horizon, or, better,
with the horizon itself playing the role of Bob,
and Alice outside, something very peculiar happens \cite{Danielson:2022tdw}:
Bob (the horizon) is aware of the superposition 
but, since no information can come from Bob to Alice,
she will never know about this and has then all the time she wants to do the recombination of $A$ as smoothly as possible,
in principle then avoiding any photon emission.
Will she get interference fringes in the end? 

At variance with what we have seen before,
we have no hope now that,
when Alice recombines adiabatically,
the two time evolutions on $B$'s side go also
to recombine, since viewed from Alice's system, an infinite time is needed for any news from her side to reach the horizon,
and then there is
no way to remove the information present on the horizon in due time.
Thus,
contrary to the ``basic'' gedanken experiment described above,
circumstances demand here that something happens
which breaks the coherence of $A$ even when the superposition
is simply held in place.

In \cite{Danielson:2022tdw} the proposal has been made that, what causes this, is the emission of one or more entangling photons by $A$
of the sort described above for the case of the gedanken experiment without horizon, but now, due to the presence of the horizon
(and its quantum vacuum,
the Unruh vacuum \cite{Unruh:1976db}
if the black hole
comes from gravitational collapse),
during the static separation phase.
These photons 
correspond to the field sourced by the difference
of the currents in the two superposed positions,
i.e., effectively, from a dipole with that charge and
separation.
They eventually go past the horizon bringing
the information of the branch
and some amount of energy
to the black hole.

Now, depending on the circumstances,
this flux of entangling photons can be shown to bring
a very small, even vanishing, amount of energy,
yet an additional finite entropy 
$\delta S = \ln 2$ 
inside \cite{Kudler-Flam:2025yur}.
This poses a challenge: 
Alice's particle evolves indeed
from a pure state
(zero entropy)
to a maximally mixed one (entropy of the two-dimensional system $\ln 2$)
while this same amount of entropy is transferred to the
black hole.
This might not be a problem in principle
for the generalised second law \cite{Bekenstein:1974ax},
which asserts that the sum of the black hole's and environment's 
entropy cannot decrease, 
as the balance can match anyway.
But, if entropy with no energy is deposited into the black hole,
a possible tension arises
with the holographic bound to entropy,
namely the statement that no more than $\mathscr{A}/4$ entropy (in Planck units)
can be packed in a region whose boundary has area $\mathscr{A}$ 
\cite{tHooft:1993dmi, Susskind:1994vu},
and also with
the Clausius relation for black holes
\cite{Jacobson:1995ab},
$\delta M = T_H \, \delta S_{bh}$ (with $\delta M$ 
and $\delta S_{bh}$ the changes of mass and entropy of the black hole and $T_H$
the Hawking temperature \cite{Hawking:1975vcx}),
in which we would have here $\delta M = 0$
while 
$\delta S_{bh} \ge \delta S \ne 0$ (with the inequality from the generalized second law).

It has been pointed out in \cite{Kudler-Flam:2025yur} that in a quantum information 
setting a specific protocol can be considered that, when applied by Alice (and Bob),
allows information to be teleported to the black hole
through entanglement between degrees of freedom inside 
and outside the horizon,
with apparently no increase of energy and entropy of the black hole.
Then this protocol shows that {\it it is indeed possible} to gain information 
inside
with no energy and entropy burden for the black hole,
and this suggests also that more general formulations of the
holographic bound or the black hole Clausius relation
might be envisaged, explicitly encompassing the entanglement. 

Aim of the present essay is to investigate
what happens generically, without specific requirements for the operations
of Alice and Bob, and to describe a mechanism,
or a complementary way to understand the process,
which leads to loss of coherence by $A$,
 at the same time
manifestly preserving the holographic entropy bound and the
black hole Clausius relation, as well as
the generalized second law.
The idea is that the entangling photons carry information about the branch,
but do not give rise to energy and entropy
flux across the horizon;
this would be so
because of the static electromagnetic field
sourced by the quantum charge:
looking from outside,
the non-vanishing field at the horizon
would correspond
to photons just hovering at the horizon;
inside, we would have radiation
arising from the dipole field there, with no flux from outside.
The two perspectives, the outside and the inside one, 
would be equivalent, complementary descriptions
of a same mechanism responsible for the loss of coherence by $A$.

\section{The decoherence effect and black hole
thermodynamics}

Our starting point is the observation that, if we add to the
description of the phenomenon the most basic element of quantumness
for the geometry we can think of, namely the existence of a limit
length and of the ensuing 
\cite{Padmanabhan:2019art, Chakraborty:2019vki}
quantum of area, 
this apparently kills the process
(at least for area quanta of order of $l_p^2$ or larger,
with $l_p$ the Planck length) if meant in terms of (soft) photons
going inside \cite{Fahn:2025fxlLet,Fahn:2025jxhArt}.
By conservation of energy, indeed, the area gap induces an energy
threshold for the process of absorption of a mode to happen,
cutting out the inside-going modes of lowest energy,
and,
as it turns out, this inhibits
the increase of the decoherence function with the time during which the superposition exists. In this case, there is
no clash with the complementarity principle anyway,
for no information about the branch is brought inside
and then Alice is allowed to recombine coherently.

We notice that the existence of an area quantum 
is not something one adds by hand to the thermodynamic description
of black holes.
Rather, it is something connatural to the white hole nature
of the past horizon with ensuing Hawking radiation,  
which is the key for the thermodynamic description,
and which is also the basic ingredient of the effect
found in \cite{Danielson:2022tdw}, as mentioned.
It is clear then that there arises something of a tension between two
different, apparently incompatible direct consequences of a same
underlying root.

To proceed, we shall consider
the explicit expression for
the total energy associated
to the entangling photons as can be calculated starting e.g. from the
results in \cite{Gralla:2023oya}.
To this aim, first of all let us consider that 
these photons are ``entangling'' in  the sense that
they are related to the decoherence $\cal D$, 
and the decoherence functional $D$,
of the superposition
by 
\begin{equation}\label{decoherence}
    \mathcal{D} 
= 1 - e^{-\frac{1}{2} D}
= 1 - |\langle A_L | A_R \rangle |
= 1 - e^{-\frac{1}{2} \langle N \rangle},
\end{equation}
that is, the average number of emitted entangling photons $\langle N \rangle$ plays exactly the role 
of the decoherence functional
(here $A_{L,R}$ are the fields sourced by the charge in the two
superposed positions, with spatial indices suppressed for ease of notation).

Next, from \cite{Gralla:2023oya}, the number $\langle N\rangle$ of entangling
photons can be written as
\begin{eqnarray}\label{number_entangling}
\nonumber
\langle N \rangle 
&=& \frac{1}{4\pi^2} \int dS \int_0^{\omega_c} d\omega\; \frac{\omega}{2\pi} \left| \tilde{A}_A^{(d)} \right|^2 \: \coth\left(\frac{\pi\omega}{\kappa}\right)\\
&=& C \frac{2}{\pi} \int_0^{\omega_c} \frac{d\omega}{\omega} \; \sin^2\left(\frac{\omega T}{2}\right) \: \coth\left(\frac{\pi\omega}{\kappa}\right)
\approx C \, \kappa \, T
\end{eqnarray}
for the time $T$
the superposition has been kept open large
(we assume to be quite far from the horizon,
and then $T$ can be equally thought as Alice's proper time
and asymptotic time), with $\kappa$ the surface gravity of the black hole, and $dS$ denotes the integration over the bifurcation surface.
${\tilde A_A^{(d)}} = \int_{-\infty}^\infty du\; A_A^{(d)} \: e^{-i\omega u}$
is the Fourier transform of the dipole field $A_A^{(d)}$
at the horizon
with respect to Killing time,
and
\begin{eqnarray}\label{C_defined}
C = \frac{1}{4\pi^2} \int dS \; \left| \tilde{A}_A^{(d)} \right|^2
\end{eqnarray}
incorporates then the effect of the field.
The notation in \eqref{number_entangling} and \eqref{C_defined} is to be understood as $\left|{\tilde A_A^{(d)}}\right|^2 := q^{AB} \, \overline{\tilde{A}}_A^{(d)} \, \tilde{A}_B^{(d)}$
where
$q_{AB}$ is the induced metric on a spatial cross-section of the horizon and an overbar denotes complex conjugation.
The field is assumed to be a top-hat function of time with
width $T$ 
and height $\hat A_A^{(d)}$,
which leads to
${\tilde A_A^{(d)}} = \hat{A}_A^{(d)} \, \frac{2}{\omega} \, \sin\left(\frac{\omega T}{2}\right)$.
$\omega_c$~in the integrals above is a cut frequency which marks
the region which is significant in the integration, the remaining part
being negligible as an effect both of the interference between 
the various modes and of a smooth-enough way the superposition is 
really opened and closed (see e.g. \cite{Fahn:2025jxhArt} for a more detailed discussion).
Its value obeys $\omega_c T \ll 1$ and $\omega_c \ll \kappa$.

From (\ref{decoherence}) we see that $\langle N \rangle = 1$ 
gives already sizeable decoherence.
Let us take $T_1$ such that $1~=~ C \kappa T_1$,
in practice regarding the time $T_1$ to have
${\cal O}(1)$ photons as the decoherence time.
We can then take as characteristic absorbed energy $\Delta E$ the energy corresponding
to $\langle N \rangle \approx 1$, that is to $T_1$.
From (\ref{number_entangling}), it can be computed as
$
\Delta E
= \int_0^{\omega_c} \omega \, n_\omega \, d\omega
$,
with
$
n_\omega = n_\omega(T_1) 
= \frac{2}{\pi} C \, \frac{1}{\omega} \, \sin^2\left(\frac{\omega T_1}{2}\right) 
\, \coth\left(\frac{\pi \omega}{\kappa}\right)
$.
This gives
\begin{eqnarray}\label{energy}
\Delta E
&=& \frac{2}{\pi^2} \, C \, \kappa \int_0^{\frac{1}{2} \omega_c T_1}
\frac{\sin^2 x}{x} \, dx
= \frac{2}{\pi^2} \, C \, \kappa \, \Big(\ln{\frac{\omega_c T_1}{2}}
- \text{Ci}(\omega_c T_1) + \gamma + \ln 2 \Big),
\end{eqnarray}
with $\gamma$ the Euler--Mascheroni constant, Ci the cosine integral function
and having chosen $x = \frac{\omega T_1}{2}$
as integration variable.
As mentioned, this energy 
can be very small, even almost vanishing if Alice is
far away from the horizon.

\begin{figure}[h]
  \centering
  \begin{minipage}{0.6\textwidth}
    \includegraphics[width=\linewidth]{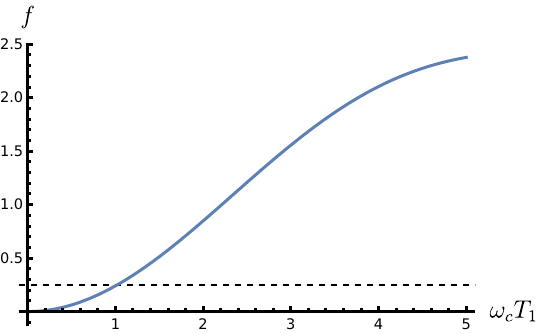}
  \end{minipage}
  \begin{minipage}{0.35\textwidth}
    \captionof{figure}{\doublespacing Plot of the term in brackets in equation \eqref{energy}. For values of $\omega_c T_1 \leq 1$ it can be estimated from above as $f\approx 0.25$ (dashed line).}\label{fig}
  \end{minipage}
\end{figure}

Indeed, in figure \ref{fig} the expression 
between round brackets in (\ref{energy}) is plotted
as a function $f$ of $\omega_c T_1$.
We see that even allowing $\omega_c T_1$ to be as large
as $\omega_c T_1 = 1$, we have $f(\omega_c T_1 = 1) \equiv f_1 \approx 0.25$,
and then
\begin{eqnarray}\label{energy_number}
\Delta E \,\, \le \,\, \frac{2}{\pi^2} \, f_1 \, C \, \kappa \,\, \approx \,\, 0.05 \, C \, \kappa.
\end{eqnarray}
Now, focusing on a Schwarzschild black hole,
and taking from \cite{Gralla:2023oya} the expression for $C$
at large distances $r_0$ from the black hole,
namely
\begin{eqnarray}\label{C_far_away}
C \,\, =  \,\,
\frac{8}{3 \pi^2} \,
\frac{q^2 d^2 M^4}{\epsilon_0 \, r_0^6} + {\cal O}\Big(\frac{1}{r_0^7}\Big),
\end{eqnarray}
we can compare this energy with what one would need to have
the process of absorption
to comply with
the black hole Clausius relation,
namely
\begin{eqnarray}\label{Clausius_BH}
\delta E_{bh} \,\, = \,\, T_H \, \delta S_{bh}
\,\, \ge \,\, T_H \, \delta S \,\, = \,\,
\frac{\ln 2}{2\pi} \, \kappa,
\end{eqnarray}
with obvious meaning of
the notation, and with the inequality coming
from the generalized second law.

To do the comparison 
between the two energies in
(\ref{energy_number}) and (\ref{Clausius_BH}),
we need
$C$ dimensionless,
i.e.,
$C_\text{d'less} 
= \frac{8}{3 \pi^2} \, 
\frac{q^2 d^2 M^4}{\epsilon_0 r_0^6} \frac{G^4}{\hbar \,c^9}$,
which, 
were the Sun a black hole, $M = M_\odot$,
the particle at Sun-Earth distance,
$r_0 = 1 \, \text{a.u.}$,
and taking $q = e$ the electron charge and $d = 1\, \text{m}$,
gives
$C_\text{d'less} \approx 
1.05 \cdot 10^{-56}$
and thus, from (\ref{energy_number}),
$\Delta E \le 0.5 \cdot 10^{-57} \, \kappa$.
The number $0.5 \cdot 10^{-57}$ 
has to be compared with 
$\frac{\ln2}{2\pi} \approx 0.11$.
It is clearly 
unthinkably smaller: no hope to have the Clausius relation satisfied.
By contrast, we notice that
a quantum of area $\mathscr{A}_0$ not lower than  $l_p^2$, say  
$\mathscr{A}_0 = \alpha \, l_p^2$ with $\alpha \gtrsim 1$, is like a perfect fit for relation (\ref{Clausius_BH}), 
as it might be expected of course by
the overall consistency of
the thermodynamic description
of black holes.
Indeed,
from $\mathscr{A} = 16 \pi M^2$
and $\delta M = \frac{1}{32\pi M} \, 
\delta \mathscr{A} = \frac{1}{8\pi} \, \kappa \, 
\delta \mathscr{A}$
for our Schwarzschild black hole,
the energy corresponding
to one single area jump
is $\delta E_{bh} = \frac{\alpha}{8\pi} \, \kappa$, which satisfies (\ref{Clausius_BH})
for $\alpha \, \ge \, 4 \ln 2 \, \approx \, 2.8$.

At this point, one might think however that the energy of non-entangling
photons should also be considered, 
namely photons associated
with the presence
of a charge there, but that carry no information about the branch.
Indeed, in the frequency range relevant for decoherence
they are indistinguishable from the photons that are instead entangling,
and their contribution should be added to the energy associated
to the process.
From (\ref{energy}) 
one can readily show that
even their contribution can be made vanishingly small.
For the new $C_\text{monopole}$ from (\ref{C_defined})
with the field of the charge replacing
the field of the dipole,
one finds
\begin{equation}
    C_\text{monopole} 
= \frac{2}{3\pi^2} 
\frac{q^2 \, M^4}{\epsilon_0 \, r_0^4} 
+ {\cal O}\Big(\frac{1}{r_0^5}\Big)
= \frac{1}{4} \, C \, \frac{r_0^2}{d^2} 
+ {\cal O}\Big(\frac{1}{r_0^5}\Big),
\end{equation}
and, coming back to our example of $M = M_\odot$ and
$r_0 = 1 \, \text{a.u.} \approx 1.5 \cdot 10^{11} \, \text{m}$
we see that even if an electron charge is
in superposition with separation as small as
conceivably possible, say $d = 10^{-10} \, \text{m}$,
yet the compensation factor $r_0^2/d^2$ is not large
enough to give for the corresponding energy something 
not irremediably small.

\section{A flux across the horizon?}

Apparently there is no way out then; we have to accept
that the energy associated with the process of loss of coherence
by the horizon can be vanishingly small, this appearing in generic circumstances hardly
reconcilable with 
some basic aspects of black hole
thermodynamics, including
the existence of a quantum of area, as described above.
Precisely this sort of stumbling block, however,
might offer the key to resolve the issue.
The situation suggests indeed that the process
of loss of coherence by $A$ comes perhaps with no need at all for energy
to cross the horizon, this then allowing to evade the constraints posed
by conservation of energy joined to the existence of an area gap.

The simple fact that we always get that same information content
$\delta S = \ln 2$ for the particle -- no matter if
the energy purportedly crossing the horizon is very small,
even vanishing in the limit --
might be taken
as a hint that the process is not necessarily tied to energy
{\it crossing} the horizon.
One might think of the Coulomb field 
effectively sourced 
by the dipole
as energy hovering at the horizon,
corresponding to entangling photons trying, so to speak,
to run away not to be caught by the
black hole but in so doing remaining
static in the spacetime, 
these photons
behaving pretty much the same way as the null rays which generate the horizon.

Their energy
would be given by (\ref{energy}).
In other words, that same formula (\ref{energy}) for the energy
applies as it is,
but now with energy meaning the energy of entangling photons
hovering at the horizon rather than flux of energy of entangling
photons across the horizon.
The smaller their energy $E$ is, the larger is of course their wavelength,
and this implies that they can carry always that amount
of information $\delta S = \ln 2$ in agreement with the
Bekenstein bound \cite{Bekenstein:1980jp}, 
$\delta S \leq 2\pi \, R \, E$, with the size $R$
taken to be their wavelength.

This is akin to what happens to a charged particle (or
to a dipole, if we want to stay with the effective source of
entangling field we have here) uniformly accelerating in Minkowski spacetime.
According to an observer co-moving with the particle
(Alice, we would say,
which is stationary in the spacetime of the black hole)
we know there is no radiation, i.e. the energy passing
through any surface in the region that the particle can access
is zero, see \cite{boulware1980radiation}
and references therein.
We do have a field there, indeed, but no energy flux,
there is no net flow of energy from the charged particle
to the field.
We have instead radiation beyond the Rindler horizon
(which in our parallel here means beyond the black hole 
horizon),
this radiation corresponding or expressing 
the retarded field of the particle there.
Then, coming back to our circumstances, 
we have no flux of radiation from the dipole to the horizon;
we have a stationary dipole field on the horizon which
we can associate to photons hovering at the horizon
with total energy $\Delta E$ as calculated above
giving rise to an energy density 
rather than 
an energy flux, 
and we have a radiation field inside the horizon, originating from the dipole
but not
moving from it, rather, expressing the dipole field
beyond the horizon. 
No energy goes from the particle to the radiation field
inside the horizon; the latter energy is the energy
of that part of the would-be stationary dipole field
beyond the horizon.
One can draw an equivalent description
in terms of the dipole field beyond
the horizon, namely from the radiation
arising inside.

The origin of the loss of coherence would be in the fact
that the modification of the field at the horizon,
embodied now by one or more entangling photons, hovering there rather than
flowing inside,
stores in an irretrievable way, and then as a permanent record
from the standpoint of Alice,
the information
about the branch.

\section{Conclusion}

Summing up,
focusing on an electrically-charged quantum system in spatial
superposition,
we have considered 
the process of decoherence
induced by horizons,
focusing on its connections with black hole thermodynamics.
We have shown that
some tension can arise, 
essentially due to the presence of
a finite amount of entropy
associated with a flux of entangling photons
that might carry vanishing total energy.
We have then discussed that
a possible way out
is if these entangling photons,
in analogy with the description of radiation
from an accelerating charge 
in the perspective of a co-accelerating observer,
are not meant as flowing across the horizon.
Rather, they would be,
from the outside perspective,
energy density of the field without a flux, 
i.e.,
like radiation ``frozen'' at the horizon corresponding
to the static Coulomb field there,
and, viewed from inside,
as radiation arising inside the black hole
from the Coulomb field present there,
in either case
with no flux of radiation
crossing the horizon.

{\bf Acknowledgements}

This research had partial support
by the INFN grant FLAG.
The authors would like to acknowledge the contribution of the COST Action CA23130. MJF has benefited from the activities of the COST Action CA23115.


\newpage

\bibliographystyle{unsrt}
\bibliography{references}

\end{document}